\documentclass[twocolumn,prb,aps,superbib,tightenlines,floatfix,superscriptaddress,showpacs]{revtex4}
\usepackage[T1]{fontenc}
\usepackage[latin9]{inputenc}
\usepackage{amsmath}
\usepackage{graphicx}
\usepackage{amssymb}
\usepackage{hyperref}
\usepackage{cases}

\begin{document}

\title{Memcomputing: a computing paradigm to store and process information on the same physical platform}

\author{Massimiliano Di Ventra}
\email{diventra@physics.ucsd.edu} \affiliation{Department of Physics, University of California, San Diego, California 92093-0319, USA}
\author{Yuriy V. Pershin}
\email{pershin@physics.sc.edu} \affiliation{Department of Physics and Astronomy and University of South Carolina Nanocenter, University of South Carolina, Columbia, South Carolina 29208, USA}

\begin{abstract}
 In present day technology, storing and processing of information occur on physically distinct regions of space. Not only does this result in space limitations; it also translates into unwanted delays in retrieving and processing of relevant information. There is, however, a class of two-terminal passive circuit elements with memory, memristive, memcapacitive and meminductive systems -- collectively called memelements -- that perform {\it both} information processing {\it and} storing of the initial, intermediate and final computational data on the {\it same} physical platform. Importantly, the states of these memelements adjust to input signals and provide analog capabilities unavailable in standard circuit elements, resulting in adaptive circuitry, and providing analog massively-parallel computation. All these features are tantalizingly similar to those
encountered in the biological realm, thus offering new opportunities for biologically-inspired computation. Of particular importance is the fact that these memelements emerge naturally in nanoscale systems, and are therefore a consequence and a natural by-product of the continued miniaturization of electronic devices. We will discuss the various possibilities offered by memcomputing, discuss the criteria that need to be satisfied to realize this paradigm, and provide an example showing the solution of the shortest-path problem and demonstrate the healing property of the solution path.
\end{abstract}

\maketitle

\section{Introduction}

In conventional computers, information is stored in the volatile--random-access memory (RAM)--and non-volatile (e.g., hard drives, solid-state drives) memories, and it is then sequentially processed by a central processing unit (CPU). Such a mode of operation requires a significant amount of information transfer to and from the CPU and the appropriate
memories. This necessarily imposes limits on the architecture performance and scalability. For instance, in the traditional von Neumann architecture the rate at which data is transferred between the CPU and the memory units
is the true limiting factor of computational speed -- the so-called von Neumann bottleneck \cite{Backus78a}.

A possible way to partially reduce this problem is to employ parallel computing, in which the program execution is distributed over several processing cores that, in many configurations, access a physically close (``local'') memory at a faster rate than the (``non-local'') memory of the other
processors. However, while several computing cores are normally combined in modern CPUs, considerable scaling on a single workstation is achieved only with specialized units (e.g., graphic processing units, GPUs) that have been traditionally used in video games and, only recently, also in scientific computing \cite{Owens07a}.

A non-incremental change in computing performance therefore requires a paradigm shift from the traditional von Neumann architecture (or similar off-springs) to novel and efficient massively-parallel computing schemes, likely based on non-traditional electronic devices. Quantum computing has been hailed as one such scheme since its inception -- attributed to Richard Feynman \cite{Feynman1986}. The parallelism in quantum computing essentially relies on
the ability of a physical system to be in a superposition of states, thus allowing a massively-parallel solution of specific problems -- such as the integer factorization \cite{Shor1997} -- that cannot be practically handled by a classical computer.
However, despite much effort in the past two decades, a practical quantum computer able to outperform a classical computer even in its most simple operations has yet to be fabricated. One of the main issues is related to the non-unitary
evolution of quantum states when the system is in interaction with one or more environments, including the
environment (apparatus) that has to eventually ``read'' the outcome of the computation \cite{Chuang95a,Pellizzari95a}.

Can we then envision an alternative computing paradigm such that {\it i}) it is intrinsically massively-parallel, and {\it ii}) its information storing and computing units are physically the same? Our brain seems to fit such a requirement. Even though the full topology of the human brain is not yet known -- the Human Connectome Project has been launched to address precisely this issue \cite{humanconnectomeproject} -- we know that the neurons and their connections (synapses) store and process information simultaneously, and their operation is {\it collective} and {\it adaptive}. Moreover, the brain's operation -- to a certain extent -- is stable with respect to failure of some amounts of neurons. Therefore, our brain boasts an embedded
 {\it self-healing} mechanism, namely, the ability to bypass broken connections with alternative paths that self-reinforce \cite{book_synapse}.
\begin{figure*}[t]
 \begin{center}
    \includegraphics[width=14cm]{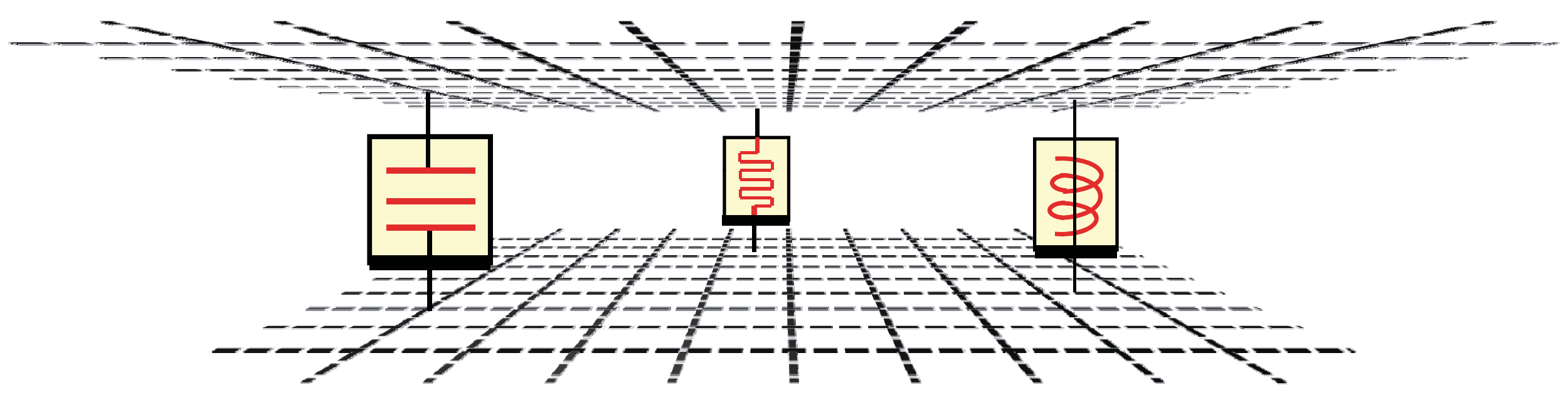}
\caption{\label{network} (Color online) Symbols of the three memory elements that we consider for memcomputing. 
Memcapacitive (left), memristive (center), and meminductive (right) systems.
}
\end{center}
\end{figure*}

In order to reproduce such features in electronic circuits, we would need elements that adapt to the incoming signal and retain information
``on demand''. Traditional transistor-based architectures would indeed fit these requirements. However, transistors are active three-terminal devices, and their operation comes at the cost of relatively high power consumption and low density.
While active elements are unlikely to be eliminated from electronics all-together, it would be desirable to keep them at a minimum,
and instead leave the storing and processing of information to passive elements, preferably with dimensions at the nanometer scale, namely
comparable to, or even smaller than their biological counterpart.

In this work we introduce the concept of {\it memcomputing}--computing using memory circuit elements (memelements)~\cite{diventra09a}--which indeed satisfies requirements {\it i)} and {\it ii)}, and does not rely on
active elements as main tools of operation. Memelements are {\it two-terminal} electronic devices whose resistance, capacitance or inductance keeps track of the system's past dynamics. They
arise naturally at the nanoscale due to the delayed response of electrons and ions in condensed matter systems subject to external time-dependent fields \cite{diventra09a}. Their general definition is quite straightforward: an $n$-th order $u$-controlled memory circuit element is defined by the set of equations~\cite{diventra09a}
\begin{eqnarray}
y(t)&=&g\left(x,u,t \right)u(t) \label{Geq1}\\ \dot{x}&=&f\left(
x,u,t\right) \label{Geq2}
\end{eqnarray}
where $f$ is a continuous $n$-dimensional vector function of internal state variables (e.g., spin polarization, ion dopants position, etc. \cite{pershin11a}), $u(t)$ is the input (voltage, charge, current, or flux), and $y(t)$ the output (the complementary constitutive variable of the voltage, charge, current, or flux). The following relations (including interchanges of $u(t)$ and $y(t)$ in each pair) then define the three different classes of elements (their symbols are shown in Fig.~\ref{Network})
\begin{equation}
\left. \begin{array}{ll}
u(t)&=\mathrm{current}\\
y(t)&=\mathrm{voltage}
\end{array}\label{memR}\right \}\rightarrow \mathrm{Memristive}\,,
\end{equation}
\begin{equation}
\left. \begin{array}{ll}
u(t)&=\mathrm{charge}\\
y(t)&=\mathrm{voltage}
\end{array}\label{memC}\right \}\rightarrow \mathrm{Memcapacitive}\,,
\end{equation}
\begin{equation}
\left. \begin{array}{ll}
u(t)&=\mathrm{current}\\
y(t)&=\mathrm{voltage}
\end{array}\label{memL}\right \}\rightarrow \mathrm{Meminductive}\,.
\end{equation}

In what follows, under {\it massively-parallel processors} we understand arrays of memelements
combined with traditional circuit components. In these structures, the information processing and storage are realized on the {\it same} platform. The computation is realized as the evolution of the system connected to external voltage (or current) sources. The collective circuit dynamics then results in an unprecedented increase~\cite{pershin11d} in computational power as we demonstrate below by applying memcomputing architectures to certain graph-theory optimization problems.

Massively-parallel analog and digital computing architectures based on memelements can be designed in several different ways using a variety of physical systems with memory~\cite{pershin11a}. All of them, however, have to satisfy general
fundamental criteria in order for this paradigm to be of value. We then formulate the six most important criteria for memcomputing. These criteria should be used as a guideline for the rational design of computing architectures. Based on these criteria, we design a memristive processor and consider certain features of its collective dynamics.

Arguably, one of the most interesting aspects of its dynamics is {\it self-reinforcement}. This feature shares a striking similarity to the collective behavior of certain biological organisms, such as ant colonies \cite{book_ants}. Therefore, the massively-parallel analog processors we discuss are ideally suited for the hardware realization of a family of related ``ant-search'' optimization algorithms. As an illustration, we consider in this paper only the solution of the shortest path problem--an important problem for several technological applications (e.g., transportation)--as well as demonstrate the healing property of the solution path. We will then conclude this paper with considerations on future directions.

\section{Memcomputing criteria}

We now formulate a few basic criteria that are both necessary and/or desirable for the implementation of computing with memory.
Some of them are similar to those introduced by DiVincenzo in the field of quantum computation~\cite{DiVincenzo00a}, which in turn are indeed similar to those required by any computing paradigm.
Others are specific to memcomputing. As briefly mentioned above, both memcomputing and quantum computation rely on massive parallelism of
information processing. The basic mechanisms of the massive parallelism, however, are quite different. While quantum computing
relies on the superposition of states, memcomputing utilizes the collective dynamics of a large number of (essentially classical) systems. Its specific criteria are then as follows.

\vspace{0.3cm}

{\it 1. Scalable massively-parallel architecture with combined information processing and storage}

Fundamentally, memcomputing is performed by an electronic circuit containing a collection of memelements (memristive, memcapacitive or meminductive systems or their combinations) that simultaneously allow for information processing and storage. By definition, memelements store information in {\it analog} form in their response characteristics (in addition to the ability of memcapacitive and meminductive systems to store information in the electric and magnetic field energies, respectively). It is expected that all, or at least a large number of memelements are involved in the parallel computation. This is the basis of the potential advantage of memcomputing over the traditional sequential one.

Combined information processing and storage is a useful feature of memelements. It simplifies the hardware design reducing the amount of components needed to realize computing functions. In this way, it becomes possible to achieve higher integration densities as well as more complex connectivities. Clearly, both factors improve the hardware functionality. For example, such a feature has been recently employed in memristive binary logic circuits, where the same memristive devices serve simultaneously as a gate and latch~\cite{strukov05a,Snider07a,borghetti10a,pershin12a}, or in memristive networks solving a maze problem~\cite{pershin11d}. Moreover, it has been shown that the performance of logic circuits improves if several types of memelements are combined together~\cite{pershin12a}. This fact should also be taken into account in memcomputing architecture design.

\vspace{0.3cm}

{\it 2. Sufficiently long information storage times}

Next, let us consider requirements that should be imposed on individual memelements. First of all, they should provide sufficiently long information storage times. At least, much longer than the calculation time. Ideally, one would use memelements with non-volatile memory storage capabilities, such as, emergent non-volatile memory cells \cite{waser07a,scott07a,sawa08a,karg08a,burr08a,pershin11a}. It is important that many of these elements are realized at the nanoscale,
and thus many of these can be incorporated on a single chip.

Additionally, it is desirable to use memelements with low power consumption and short read/write times.
Emergent non-volatile memory cells satisfy these requirements and thus are ideal candidates for memcomputing architectures. For example \cite{jo09a}, CMOS compatible nanoionic resistive switches based on amorphous-Si offer promising switching characteristics
in terms of write speed (<10 ns), endurance (>$10^5$ cycles), retention
($\sim$7 years), and scaling potential (<30 nm). Asymmetric high-endurance Ta$_2$O$_{5-x}$/TaO$_{2-x}$ bilayer structures
sustain up to $10^{12}$ write cycles \cite{lee11a}. Finally, the energy to write a bit of information into a nanoionic cell can be as small as $5\cdot 10^{-14}$J, \cite{itrs09a}, which represents a major energy saving.

\vspace{0.3cm}

{\it 3. The ability to initialize memory states}

Memelements should be initialized before the computation begins -- the initialization could be provided automatically in the case of volatile memelements, and may not be needed for memelements storing intermediate/final values of parameters. In any case, the computing device should provide a mechanism for initialization of relevant memelements. This is expected to be an easy task because, typically, memdevice response functions change between two limiting values. For instance, in the case of the most studied bipolar memristive devices, the application of a high amplitude pulse of a given polarity for a sufficiently long time interval guarantees the device switches into one of its limiting states (depending on the pulse polarity).

\vspace{0.3cm}

{\it 4. Mechanism(s) of collective dynamics, strong "memory content"}

The device architecture should provide a mechanism of collective dynamics in which the evolution of a memdevice state depends on the states of several/all other devices.
For example, in the case of memristive logic, voltages applied to a couple of memristive devices change the state of one of these depending on the state of the second one.
In order to obtain a reliable switching, we also require a strong "memory content": the device characteristics in its limiting states should be sufficiently different to
provide a significant influence on other memelements.

\vspace{0.3cm}

{\it 5.  The ability to read the final result (from relevant memelements)}

Once the computation is performed, the read out of the information needs to be done (preferably) without modifying the states of the single memelements. This can be accomplished if we choose a reading input $u(t)$ such that the device state (described by the state equation \ref{Geq2}) stays constant or varies little. Generally, it is not a problem in a system based on memelements with threshold \cite{pershin11a}. This can also be done with threshold-less devices, however by appropriately tailoring the reading input $u(t)$ to minimize the state change.

\vspace{0.3cm}

{\it 6. Robustness against small imperfections and noise}

Insensitivity to small imperfections of computer components or small "damages" of the architecture should be an
 essential feature of (most) memcomputing architectures. Indeed, small imperfections are always introduced during the fabrication of each element and are therefore unavoidable. The computer architecture should be thus robust with respect to such imperfections. "Damages" are more serious deviations from the ideal computing structure that can be introduced during the fabrication and/or use of the computing device. In the human brain, for example, although multiple neurons die each day, the overall brain functionality is not influenced for many years. Likewise, it is reasonable to require that the operation of memelements networks is not sensitive to relatively small "damages".

\begin{figure}[tb]
 \begin{center}
    \includegraphics[width=5cm]{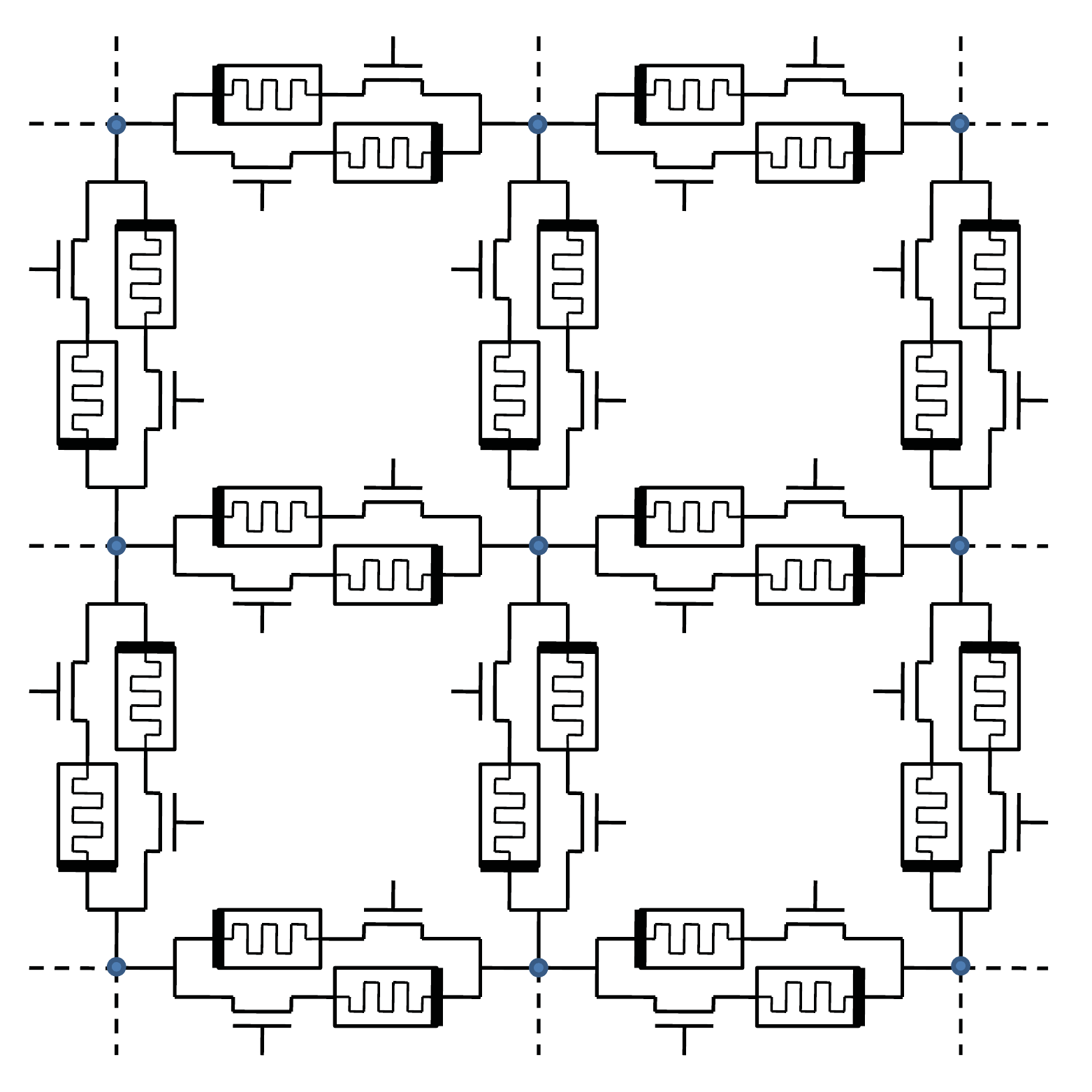}
\caption{\label{fig1} Memristive processor consisting of a network of memristive elements in which
each grid point is attached to several basic units. Each basic unit involves
two memristive devices connected symmetrically (in-parallel) and
two switches (field-effect transistors). The switches provide access to individual memristive devices, while
in-parallel connection symmetrizes the response of bipolar memristive elements.
}
\end{center}
\end{figure}

\section{Memcomputing schemes}

There are several schemes that have been recently suggested that satisfy all or some of the above criteria.
These schemes include neuromorphic computing with memristive synapses \cite{snider08a,Barranco09b,Afifi09a,pershin10c,jo10a,Ebong12a,pershin12a}, massively-parallel computing with memristive networks \cite{pershin11d}, logic with memory circuit elements \cite{strukov05a,Lehtonen09a,borghetti10a,pershin12a}, and memristive cellular automata \cite{Itoh10a}.
For instance, in the work of Ref. \onlinecite{pershin11d} a memristive network has been used to solve a popular optimization problem,
namely the maze problem. Criterion {\it 1} is fully satisfied by that network: the solution of the maze is done in an analog massively-parallel fashion, and it is locally stored in the system for essentially an unlimited
time (criterion {\it 2}). The network can also be initialized easily as explained in Ref. \onlinecite{pershin11d}, thus
satisfying criterion {\it 3}. The dynamics of the system is collective, and the difference between the low-resistance state and the high-resistance state was chosen in order to easily perturb all memelements in the system
(criterion {\it 4}). Although not explicitly discussed in that work, criterion {\it 5} can be easily
accomplished with an appropriate choice of memelements. Finally, criterion {\it 6} was naturally built in the problem: any change of topology of the network--and consequent emergence of new maze solution(s)--could be handled
effortlessly. Similar considerations would apply to the other schemes proposed in the literature.

We note at this point that although they are not extensively studied yet, memcapacitors and meminductors \cite{diventra09a} can also be used in the above memcomputing schemes by replacing memristors, albeit in a modified form. Since memcapacitors and meminductors may in principle be constructed to consume little or virtually no energy, their use in memcomputing is potentially
energetically more efficient than the use of memristors. An important milestone in this field would be the demonstration of a memcomputing device with computing capabilities and power consumption comparable to (or better than) those of the human brain.

\begin{figure*}[bt]
 \begin{center}
  \centerline{
    \quad \quad \quad \quad \quad
    \mbox{\includegraphics[width=7.00cm]{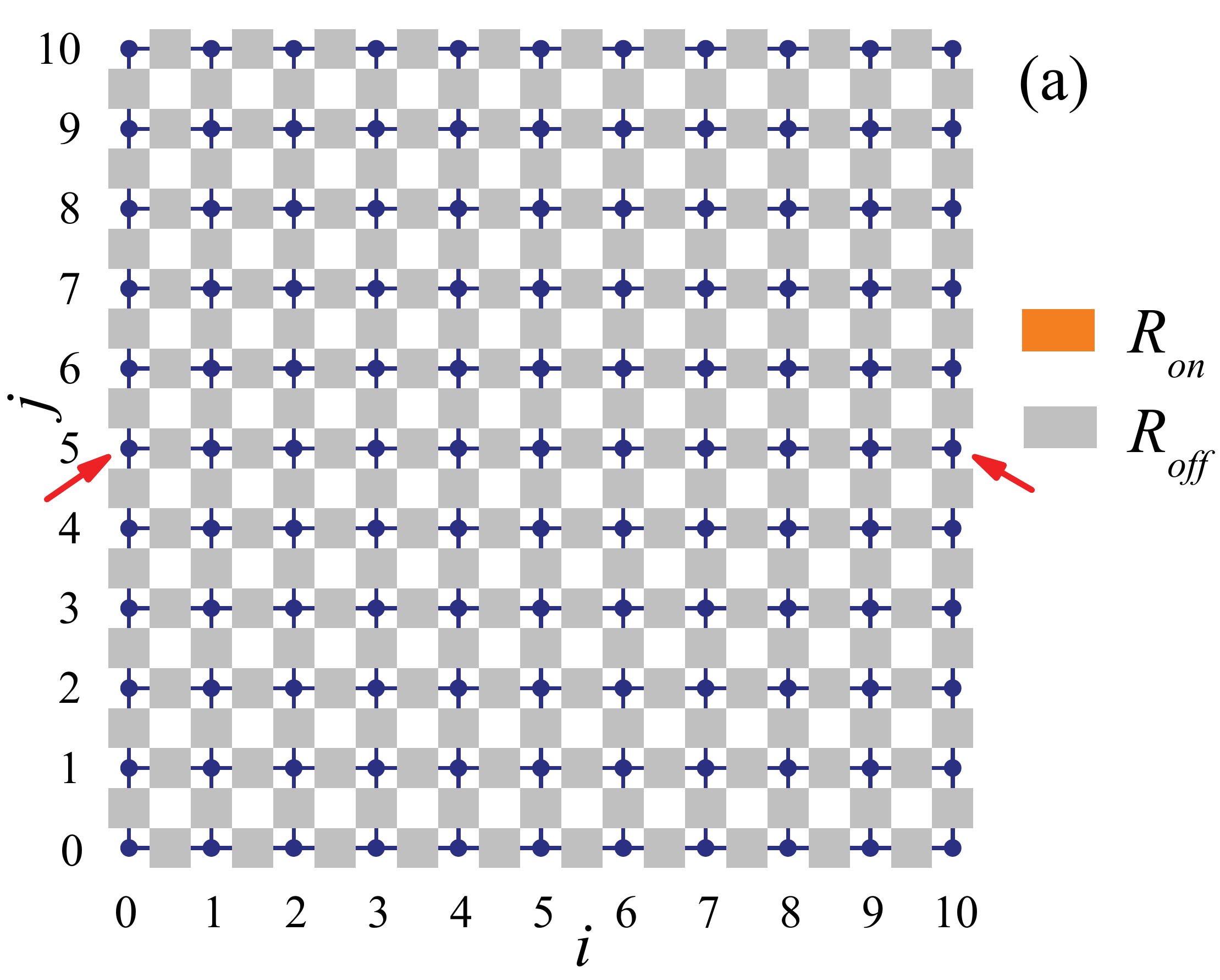}}
    \quad \quad
    \mbox{\includegraphics[width=7.00cm]{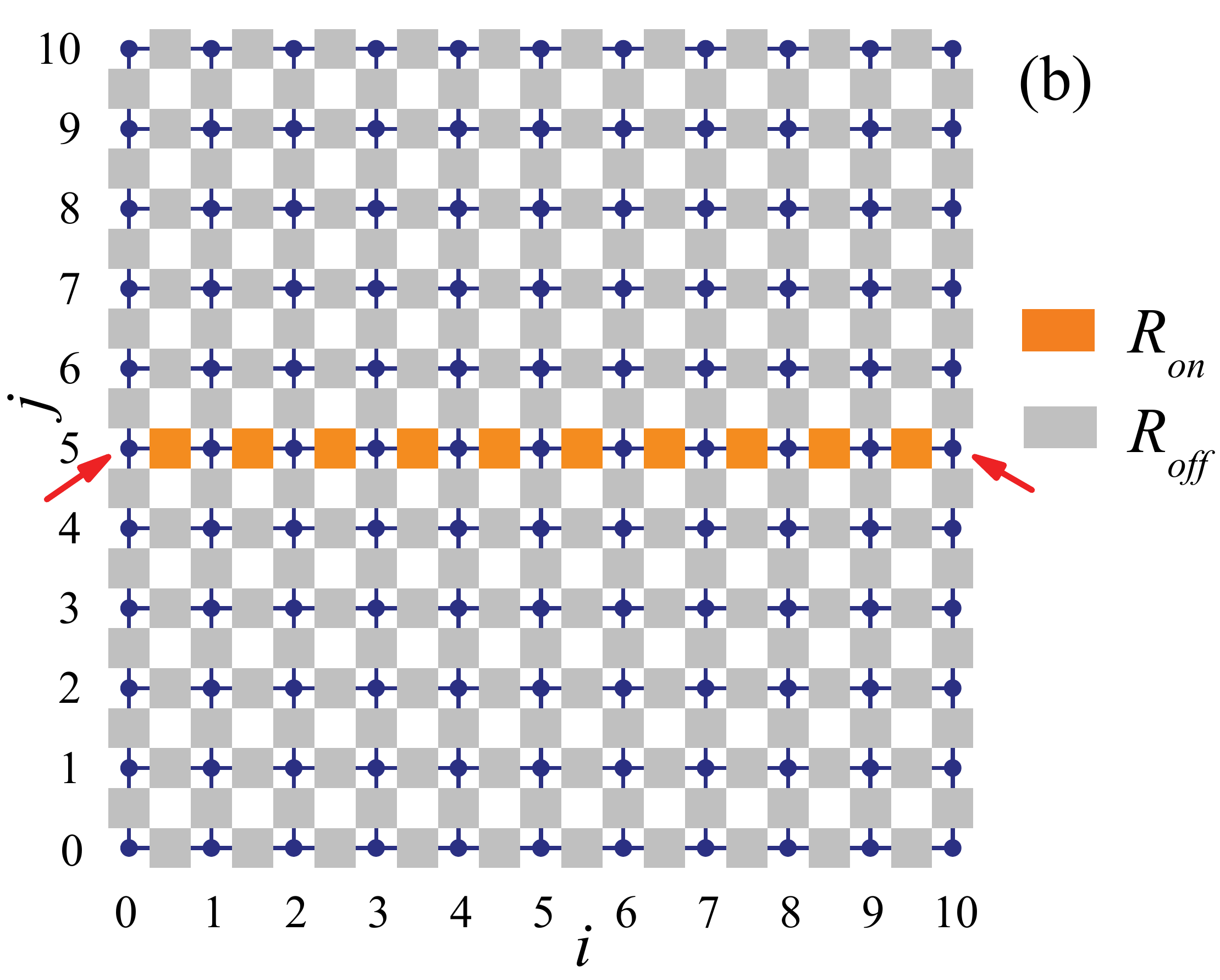}}
  }
  \centerline{
    \mbox{\includegraphics[width=7.00cm]{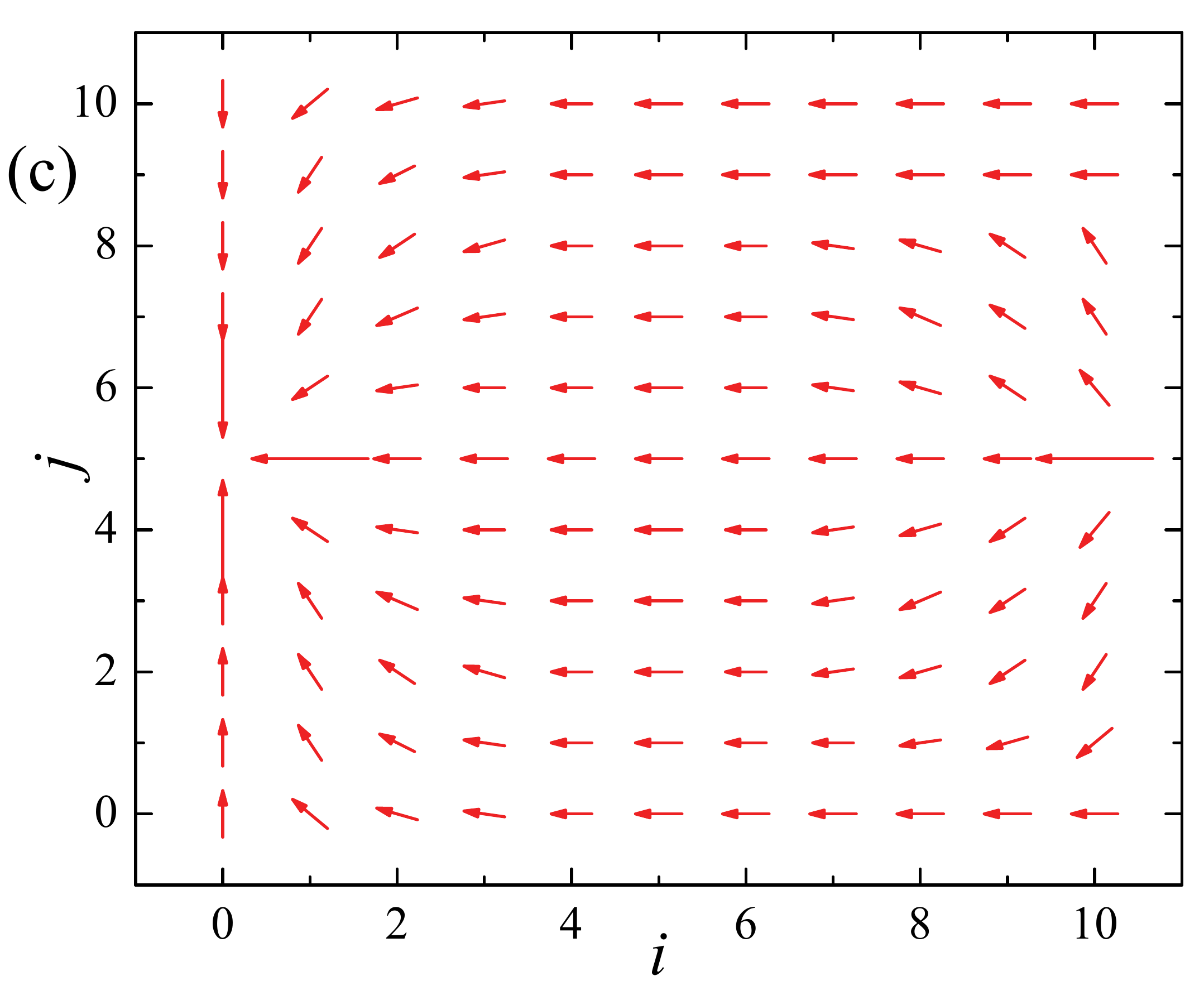}}
    \quad \quad
    \mbox{\includegraphics[width=7.00cm]{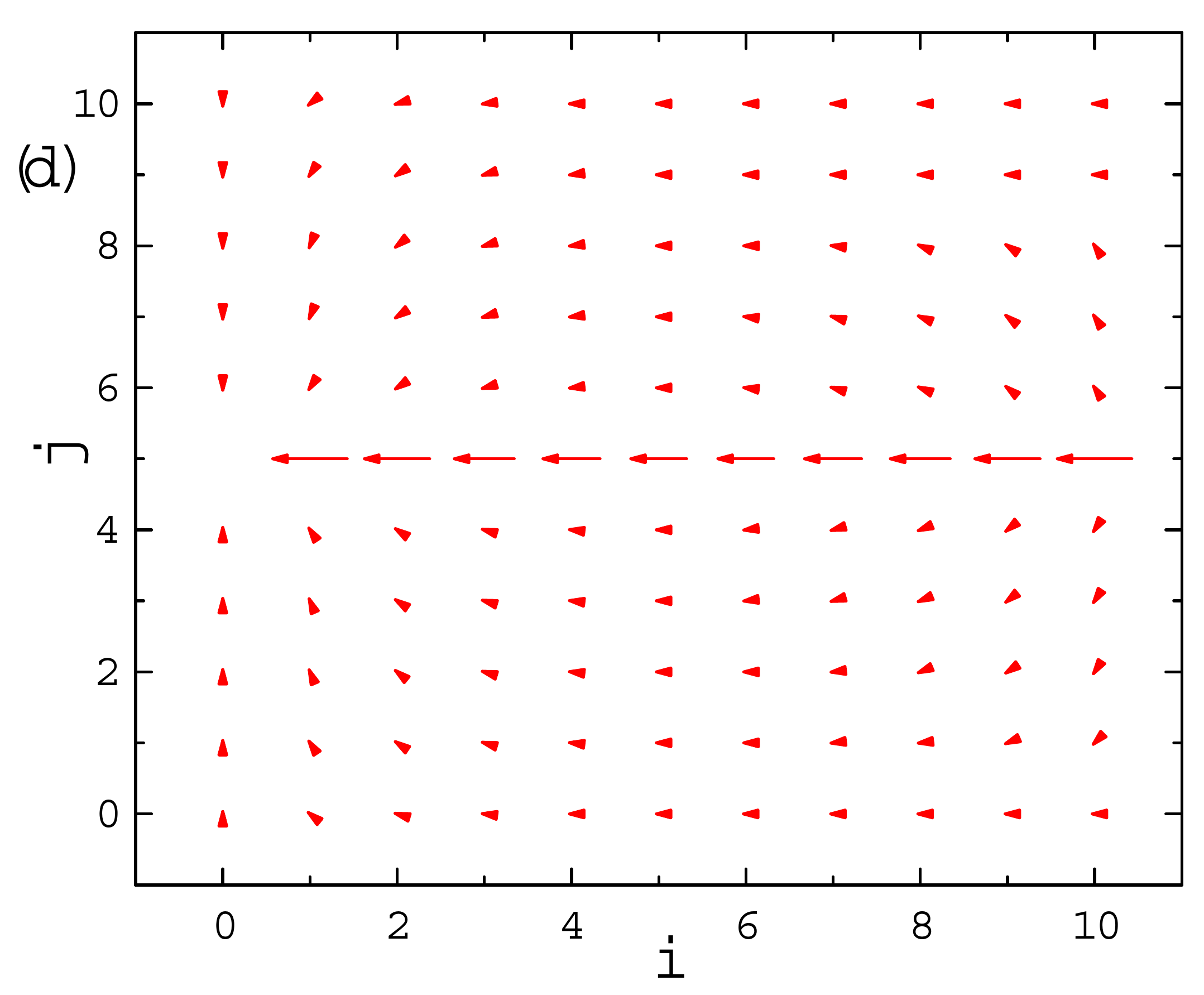}}
  }
 \caption{\label{fig2} (Color online). Solution of the shortest path problem for the pair of nodes indicated by red arrows in (a) and (b) in a an 11$\times$11 memristive network. (a) Initial and (b) final states of the memristive network. Here, the memristance of each basic unit (involving two memristive devices) is represented by a color. Distributions of electron current corresponding to (a) and (b) are shown in (c) and (d), respectively. See
 Methods for details of all calculations. }
\end{center}
\end{figure*}

\section{Application of memristive networks to the shortest-path problem}

We now provide an explicit example of computing with memelements -- specifically with memristive networks -- in order to exemplify even further the criteria given above, and the possibilities offered by this paradigm. Details
of the specific memristive systems used and the simulation details can be found in the Methods section. The memristive network (processor) consists of
a square array of grid points connected by basic units involving two bipolar memristive elements and two switches (see Fig. \ref{fig1}). Although not strictly necessary, in this example we have chosen the design of
each basic unit to be symmetric so as to conveniently provide independency of the circuit operation on the sign of the applied voltage. The switches are introduced for two functions: {\it i}) to provide independent access to each individual memristive device, and {\it ii}) to define the network topology, if needed (see Ref. \onlinecite{pershin11d} for an example). Thus, the architecture of Fig. \ref{fig1} provides access to each individual memristive device for the purpose of initialization and reading of the calculation result. The calculation consists in the evolution of the network state--defined as the collection of states of all memristive devices--when an appropriate pulse sequence is applied to a sub-set of grid points.

In order to demonstrate the main principles of memcomputing, let us consider the solution of the shortest-path problem along a symmetry direction of the memristive processor \footnote{Different grid and connectivity patterns are needed to solve the shortest path problem in more general cases. Such a study is out of scope of this publication that focuses on general criteria for memcomputing.}
Specifically, we would like to find the shortest path between two pre-selected nodes (shown by red arrows in Fig. \ref{fig2} (a) or (b)) in the network.
The algorithm used for this calculation is the following:

{\it Initialization stage}: all memristive devices are pre-initialized into the "OFF" (high-resistance) state.

{\it Calculation stage}: a voltage pulse of a suitable amplitude and duration is applied to the pair of pre-selected nodes.

{\it Calculation reading}: the shortest path is given by a sub-set of basic units in their "ON" (low-resistance) state. \footnote{The memristance of a basic unit is that of two memristive devices connected in-parallel. Note that in-parallel connection of two memristive devices can be described as a single higher-order memristive device.}

Fig. \ref{fig2}(a) shows the initial state of the network when all memristive devices are in their "OFF" states. At the initial moment of time $t=0$, a single constant amplitude voltage pulse is applied to the input/output nodes shown by red arrows in Fig. \ref{fig2}(a). The final state of the memristive network (the calculation result) is presented in Fig. \ref{fig2}(b). Clearly, two pre-selected nodes are connected by a chain of memristive devices in the "ON" state giving the shortest path problem solution. Note that Fig. \ref{fig2}(a), (b) depict the memristance of each basic unit consisting of two memristive devices.

\begin{figure}[bt]
\begin{center}
    \includegraphics[width=7.0cm]{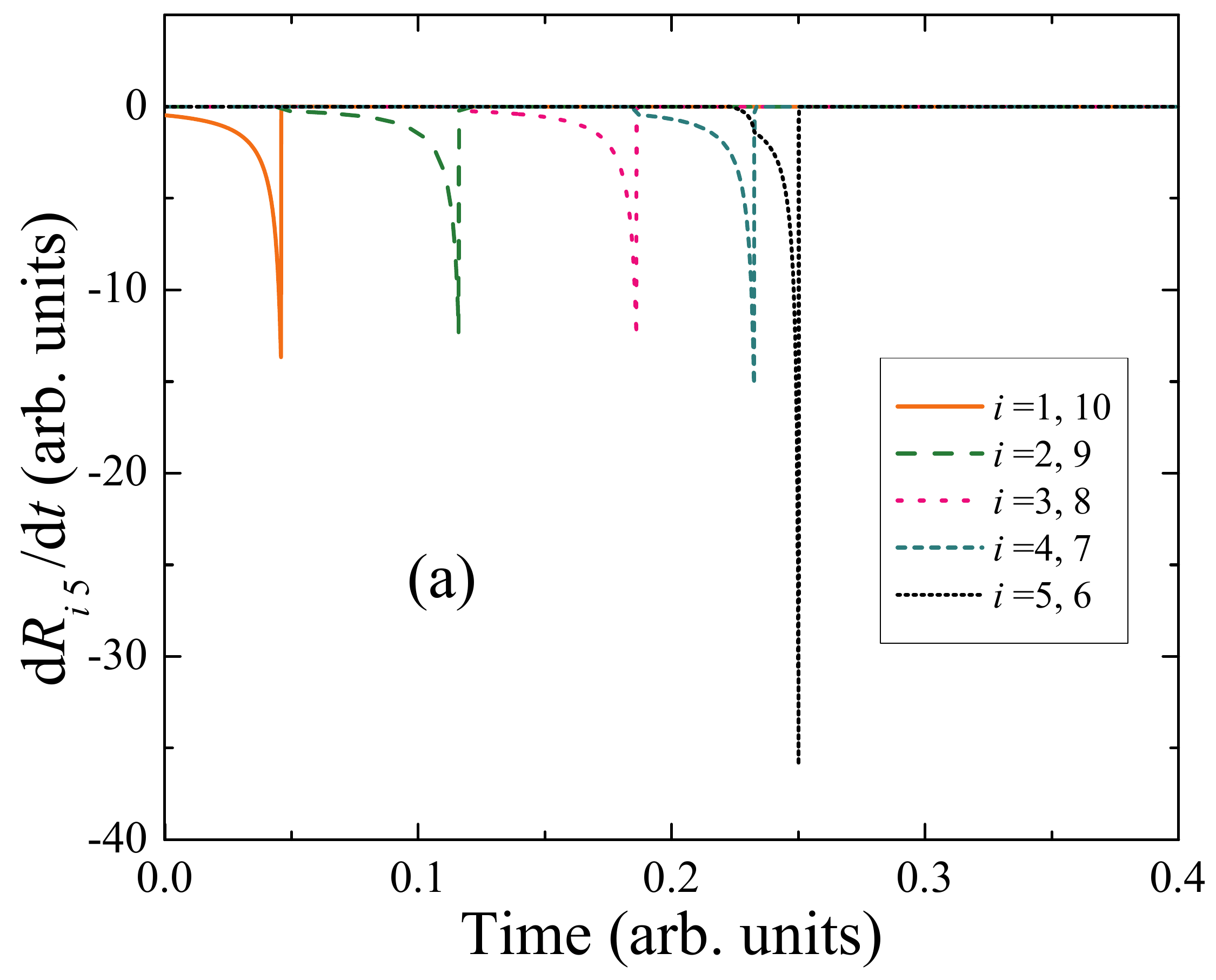}
    \includegraphics[width=7.0cm]{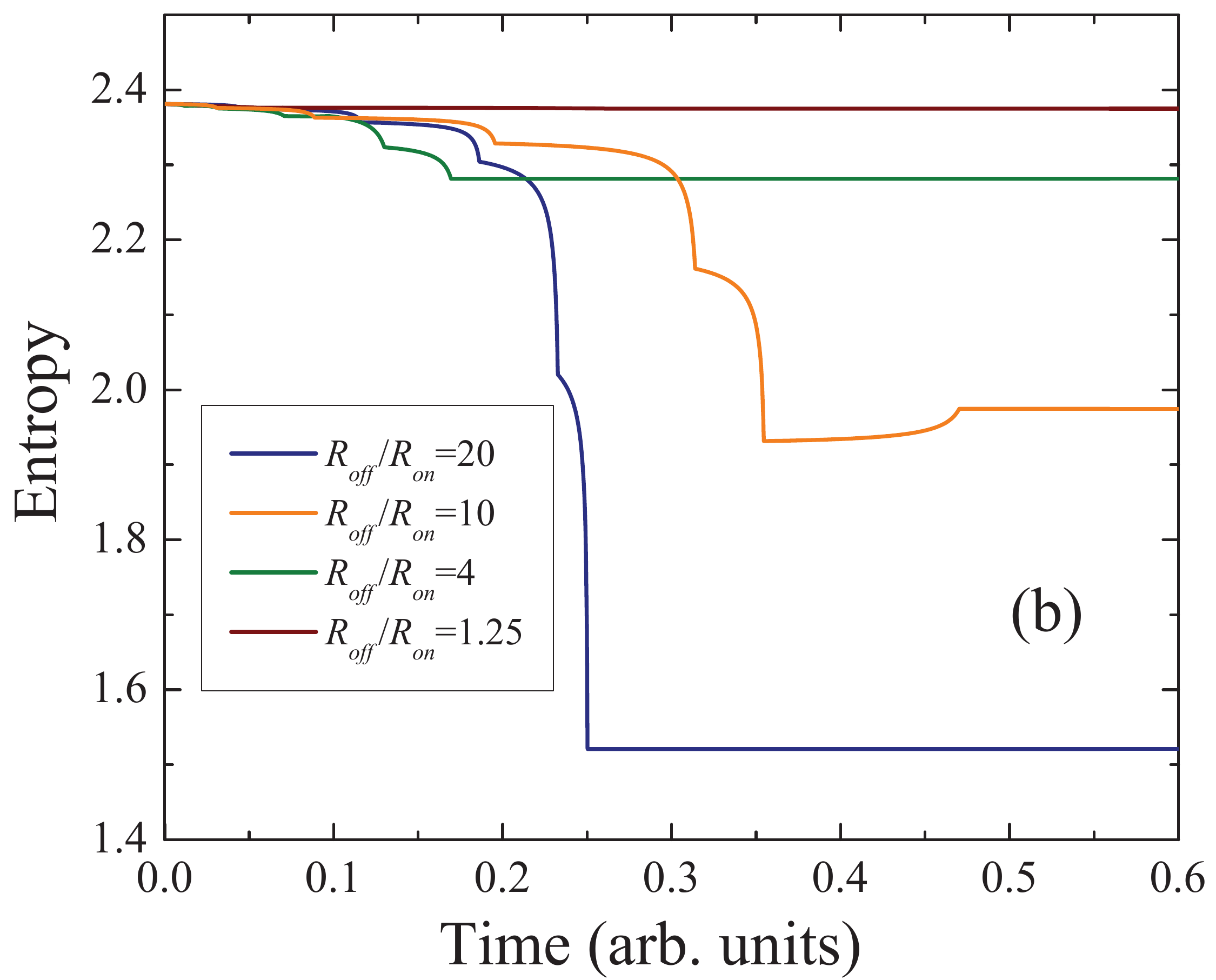}
\caption{\label{fig3}  (a) Dynamics of resistance switching within the calculation stage corresponding to the shortest-path problem solution
presented in Fig. \ref{fig2}. This plot shows that the solution emerges from both sides and propagates to the center.
(b) Network entropy as a function of time from Eq. (\ref{eq_entropy}) for networks of different memory content. A slight increase of the entropy for the $R^M_{off}/R^M_{on}=10$ curve at the time of about 0.4 (arb. units) is due to a delayed switching of four vertical units directly connected to the input/output nodes.}
\end{center}
\end{figure}

Let us now consider the network evolution--the dynamics of the calculation stage--in more details. First of all, we would like to mention a similarity between the process of computing as performed by the memristive processor and the ant-colony optimization algorithm \cite{Colorni91a,book_ants}. The latter is an adaptable algorithm inspired by the observation that ants, upon finding food, mark their trails by pheromones thus attracting other ants in order to reinforce the trail closest to their nest. A similar type of reinforcement is observed in the memristive network dynamics. Fig. \ref{fig2}(c) shows the current distribution in the network at the initial moment of time when all the memristive devices are in the "OFF" state. In this case, the current flows in multiple paths. However, since the rate of memristance change is proportional to the current, the memristance of the least resistive path will decrease faster ``attracting'' more and more current. Therefore, the current flowing through the least resistive path will reinforce this path, similarly to the trail reinforcement of the ant colony, see Fig. \ref{fig2}(d).

Moreover, it is interesting to note that the problem solution develops gradually, starting from both the pre-selected nodes. This is clearly seen in Fig. \ref{fig3}(a) that presents the rate of change of the memristances along the solution path as a function of time.

In order to quantify the system evolution even further, we define a {\it network entropy} with respect to memristances, or equivalently currents in the network. For example, by considering currents through a vertical cross section of the network at its center, we can define the network entropy as (note that in the absence of capacitive components the total current is independent of the choice of this cross section at any given time,
provided such cross section crosses the shortest path solution and spans the far upper and lower ends of the network without self-intersecting)
\begin{equation}
\sigma(t)=-\sum\limits_{i=1}^{N}\tilde I_{ij}(t)\textnormal{ln}\left( \tilde I_{ij}(t) \right),
\label{eq_entropy}
\end{equation}
where $N$ is the number of basic units connected in the horizontal direction, $j$ is the index of the row of horizontal basic units crossed by the cross-section, $\tilde I_{ij}=I_{ij}/I_{tot}$ is the normalized current through a horizontally connected basic unit, and $I_{tot}=\sum_{i=1}^{N} I_{ij}$. Fig. \ref{fig3}(b) demonstrates (with
$N=11$ and $j=5$) that the network entropy decreases in time as the computation proceeds. In statistical physics, the entropy is related to the number of states available to the system. Here, its decrease can be thus interpreted as due to the decrease in the number of paths available for the current, with a more pronounced decrease the larger the memory content in the system (as represented by the ratio $R_{off}/R_{on}$).

\begin{figure}[b]
 \begin{center}
    \includegraphics[width=7.5cm]{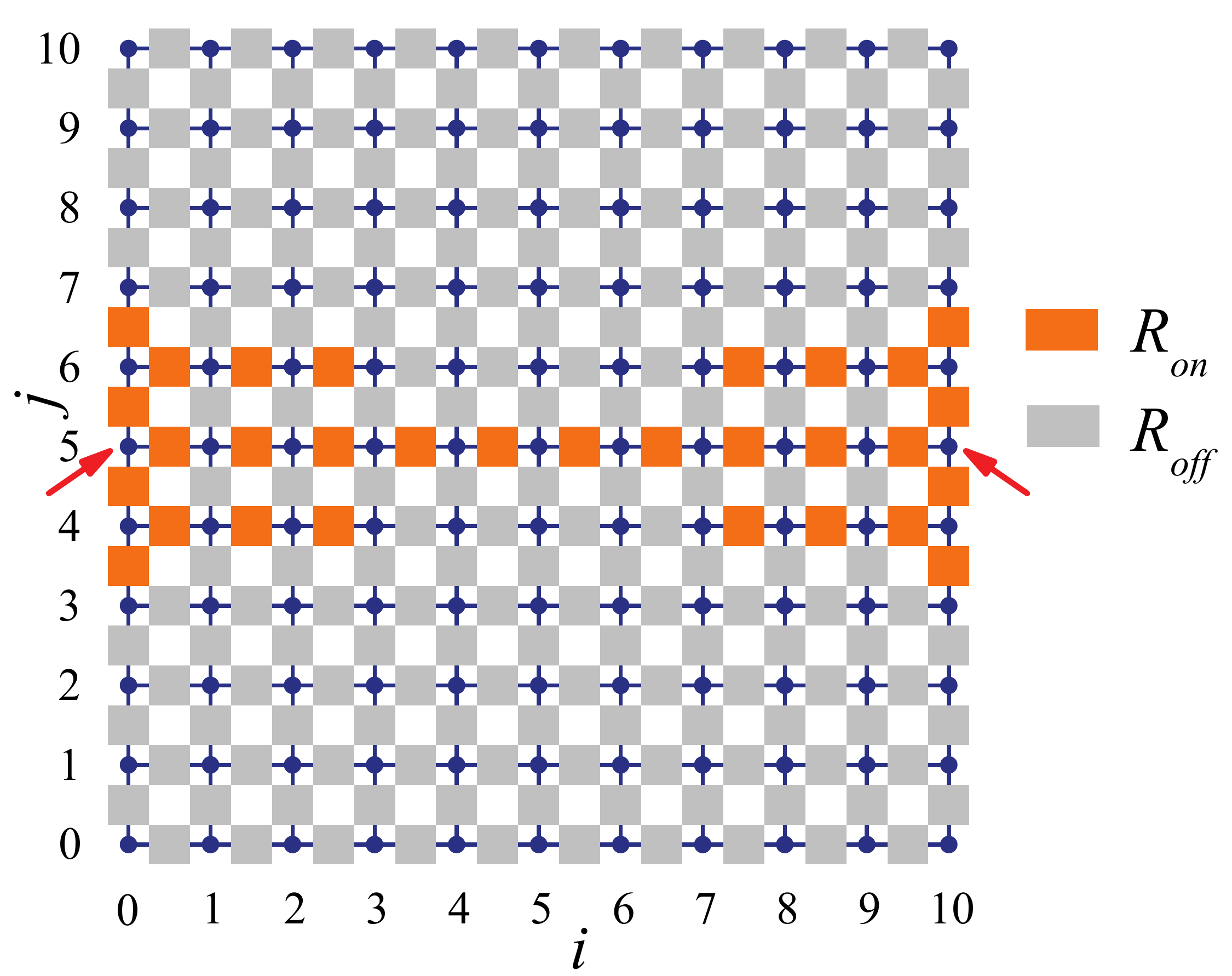}
\caption{\label{fig4} (Color online). Solution of the shortest path problem by a network of low memory content, $R^M_{off}/R^M_{on}=1.25$.}
\end{center}
\end{figure}

In order to prove this last statement, we perform a calculation similar to that presented in Fig. \ref{fig2} assuming, however, a much smaller difference between $R^M_{on}$ and $R^M_{off}$ of each memristive device ($R^M_{off}/R^M_{on}=1.25$ compared to $R^M_{off}/R^M_{on}=20$ used in the previous calculation). Fig. \ref{fig4} demonstrates that now the solution of the shortest path problem can not be found exactly at the
given bias. In fact, in addition to the switching of memristive devices directly connecting the input and output nodes, many other memristive devices are also switched into the "ON" state (see Fig. \ref{fig4}). This example demonstrates the importance of the strong "memory content" requirement (criterion {\it 4}).

\begin{figure}[bt]
 \begin{center}
    \includegraphics[width=7.5cm]{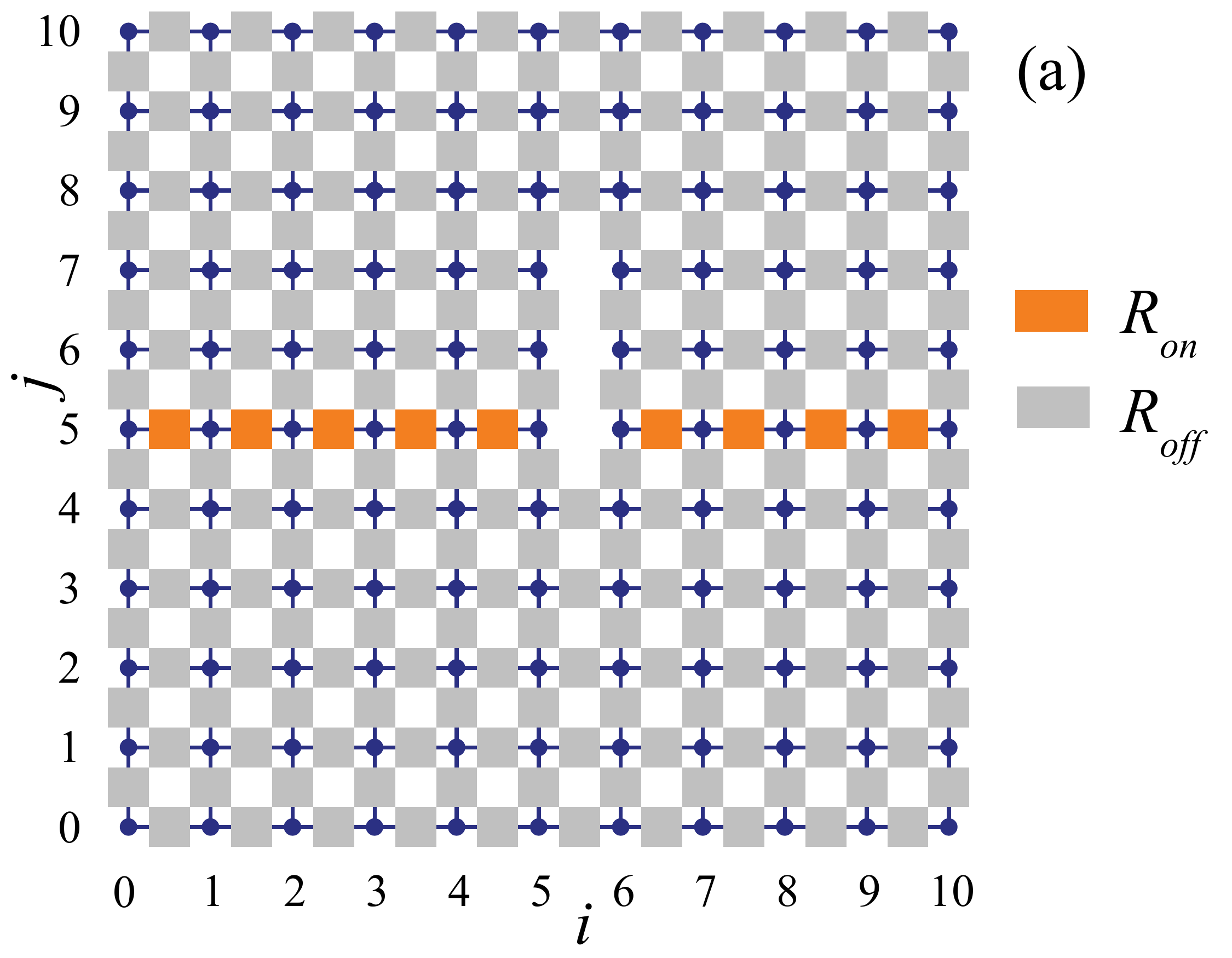}
    \includegraphics[width=7.5cm]{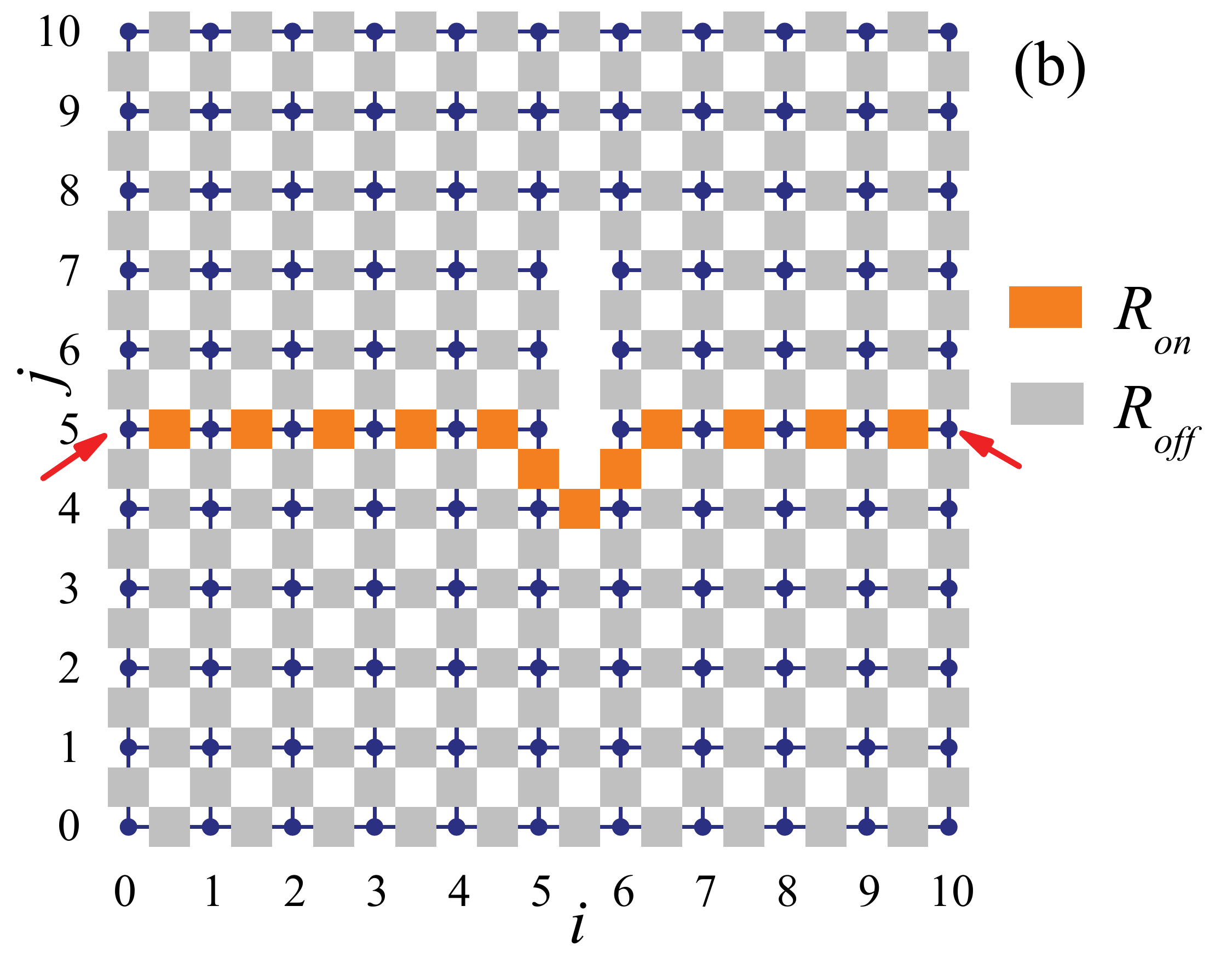}
\caption{\label{fig5} (Color online). Healing (b) of a damaged (a) solution. To heal the solution damage in (a),
a single square pulse of appropriate width and duration is applied to the input and output nodes shown by the red arrows in (b).}
\end{center}
\end{figure}

In order to analyze criterion {\it 6} in more detail, let us now damage the shortest-path problem solution shown in Fig. \ref{fig2}(a) by removing three grid points in the central part of the network as shown in Fig. \ref{fig5}(a). We note that the memristive network has a remarkable ability to {\it repair} damaged solutions -- the healing ability we have mentioned above. Indeed, this property is close to the self-healing ability that can be ascribed to systems or processes, which by nature or by design tend to correct any disturbances.

The healing of the damaged solution is performed by applying a single pulse of a certain amplitude and duration to the input and output nodes. The result presented in Fig. \ref{fig5}(b) shows that a new path connecting two pieces of the initial shortest-path solution develops below the damaged region. The three missing grid points have been removed intentionally in an asymmetric fashion in order to show that the healing occurs along the shortest possible path around the damaged region.
It is easy to understand the origin of this healing process: as soon as we switch on the pulse between the input and
the output nodes, the current will flow through all possible paths. However, the shortest one is again the one that is mostly affected, and thus reinforced during dynamics.

The stability of the shortest path problem solution to small imperfections of the system (e.g., finite width distributions of threshold current, limiting values of memristance, etc.) is evident, and therefore does not deserve a closer inspection.

\section{Conclusion}

In conclusion, we have discussed the concept of {\it memcomputing}: storing and processing of information on the same
physical platform. In particular, we have outlined the main criteria that need to be satisfied in order to realize such a paradigm
and analyzed a specific example to show the healing properties of the solution. Unlike other promising but more speculative proposals, like quantum computing, memcomputing
is already a practical reality, at least in regard to some applications, such as digital logic. It bypasses several
of the bottlenecks of present-day computing architectures and its constitutive units -- memristors, memcapacitors, and
meminductors -- are already widely available. Indeed, these elements emerge quite naturally with increasing  miniaturization of electronic devices. The computational possibilities offered by this paradigm are varied, and due to its tantalizing similarities both with some features of the brain as well as with the collective properties of colonies of living organisms, it promises to open new directions in neuromorphic architectures and biological studies.

\section{Acknowledgment}

This work has been partially supported by NSF grants No. DMR-0802830 and ECCS-1202383, and the Center for Magnetic Recording Research at UCSD.

\section*{Methods}\label{cal_details}

The numerical results presented in this paper have been obtained for a network of current-controlled bipolar memristive devices with threshold. Each memristive device is described by
\begin{equation}
V_M=R\left(x \right)I_M, \label{Icontr4}
\end{equation}
and
\begin{numcases} {\frac{\textnormal{d}x}{\textnormal{d}t}=}
0 & for $|I_M|<I_t$  \label{Icontr5}
\\
\textnormal{sgn}\left( I_M\right)\gamma\left(\left|I_M\right|-I_t\right) & for $|I_M|\geq I_t$ \;\;\;\; \label{Icontr6}
\end{numcases}
where $V_M$ and $I_M=\dot{q}(t)$ denote the (time-dependent) voltage and current across the
device, respectively; $R(x)\equiv x$ is the memristance that changes between two limiting values $R^M_{on}$ and $R^M_{off}$;
$x$ is the internal state variable; $\gamma$ is a constant describing the rate of change of memristance when
the magnitude of the electric current $I_M$ exceeds the threshold current $I_t$; and $\textnormal{sgn}$ is the sign  function.
We note that a current-controlled threshold-type memristive device model was used to describe switching in bipolar memristive devices \cite{Pickett09a}.
Moreover, many models of voltage-controlled memristive devices can be easily reformulated in the current-controlled form \cite{pershin11a}.

All numerical results were obtained for  an 11$\times$11 memristive network using in all calculations the following model parameters: $R^M_{off}=200$ Ohms, $R_{ij}(t=0)=R^M_{off}$, $\gamma=10^6$Ohms/(s$\cdot A$), $I_t=10$mA. Figs. \ref{fig2}, \ref{fig3}(a) and \ref{fig5} are obtained with $R^M_{on}=10$ Ohms and $V=6$V of applied voltage; Fig. \ref{fig3}(b) is found using $V=6,6.75,10,15.25$V for $R^M_{off}/R^M_{on}=20,10,4,1.25$ curves, respectively; Fig. \ref{fig4} is plotted with $R^M_{on}=160$ Ohms and $V=15.25$V. Note that $R^M_{on}$ and $R^M_{off}$ are
related to individual memristive devices, while $R_{on}$ and $R_{off}$ (used in Figs. \ref{fig2}, \ref{fig4} and \ref{fig5}) represent limiting values of memristance of the basic unit. While the "OFF" state of the basic unit is attained when both memristive devices are in their "OFF" states, the "ON" state of the basic unit corresponds to the "ON", "OFF" combination of single device states.
In our simulations, at each time step, the potential at all grid points is found as a solution of Kirchhoff's current law equations obtained using a sparse matrix technique. The corresponding change in the memristive states was computed using Eq. (\ref{Icontr6}). The width of the voltage pulse is selected sufficiently long to reach the steady state in each calculation.

\bibliography{memcapacitor}
\end{document}